\documentclass[aps,pra,twocolumn,superscriptaddress]{revtex4}

\usepackage{mathtools}
\usepackage{amssymb}
\usepackage{graphicx}
\usepackage{bbold}
\usepackage{hyperref}
\usepackage{xcolor}

\usepackage{xcolor}

\usepackage{soul}

\hypersetup{colorlinks=true, linkcolor=blue, urlcolor=blue, citecolor=blue}

\newcommand{\sys}{\mathcal S}
\newcommand{\env}{\mathcal E}
\newcommand{\uba}{Departamento de F\'\i sica, FCEyN, UBA, Pabell\'on 1, Ciudad Universitaria, 1428 Buenos Aires, Argentina}
\newcommand{\ifiba}{Instituto de F\'\i sica de Buenos Aires, UBA CONICET, Pabell\'on 1, Ciudad Universitaria, 1428 Buenos Aires, Argentina}
\newcommand{\unilu}{CSSM, Physics and Materials Science,
University of Luxembourg, L-1511 Luxembourg, Luxembourg}

\begin{document}
\title{Entanglement generation in quantum thermal machines}

\author{Milton Aguilar}
	\email{mil@df.uba.ar}
	\affiliation{\uba}
	\affiliation{\ifiba}
\author{Nahuel Freitas}
	\email{nahuel.freitas@uni.lu}
	\affiliation{\unilu}

\author{Juan Pablo Paz}
	\email{paz@df.uba.ar}
	\affiliation{\uba}
	\affiliation{\ifiba}

\date{\today}

\begin{abstract}
We show that in a linear quantum machine, a driven quantum
system that evolves while coupled with thermal reservoirs, entanglement
between the reservoir modes is unavoidably generated.
This phenomenon,
which occurs at sufficiently low temperatures and
is at the heart of the third law of thermodynamics, is a consequence
of a simple process: the transformation of
the  energy of the driving field into pairs of excitations in the reservoirs.
For a driving with frequency $\omega_{d}$ we
show entanglement exists between
environmental modes whose frequencies satisfy the
condition
$\omega_{i} + \omega_{j}=
\omega_{d}$. We show that  this entanglement can
persist for temperatures that can be significantly higher than
the lowest achievable ones with sideband resolved
cooling methods.
\end{abstract}

\maketitle

\section{Introduction}
Quantum thermodynamics \cite{andersVinjanampathy,kosloff,brandaoHorodeckiNg} is an emerging field whose
goal is to study the exchange of heat and
work in the quantum domain. In recent years
novel thermal machines operating at the atomic
scale  have been built using various technologies.
These include, most notably, ion traps
 \cite{eschnerMorigiSchmidt,abahRossnagelJacob,maslennikovDingHablutzel} and
superconducting qubits
\cite{karimiPekolaCampisi}.
Although a first principle description of such devices needs to be
based on quantum laws, a study of their
performance
revealed that they satisfy the
same constraints imposed by classical
thermodynamics (for example, their efficiency
is bounded by the Carnot limit) \cite{allahverdyanHovhannisyanJanzing,levyAlickiKosloff,wuSegalBrumer,ticozziViola,masanesOppenheim,wilmingGallego,skrzypczykBrunnerLinden}. Although it is
reassuring that classical results are reobtained  from a
quantum treatment, this naturally
raises a troubling
question: what is quantum in quantum thermodynamics? \cite{uzdinLevyKosloff}.
More precisely: are there
thermodynamical tasks that are possible (or impossible)
because of quantum effects?
(or: are there
quantum effects that are unavoidably associated
with thermodynamical
cycles?).
Even though, in this context, the role of quantum coherences \cite{kammerlanderAnders},
correlations \cite{sapienzaCerisolaRoncaglia}, and entanglement \cite{brunnerHuberLinden,bohrBraskHaackBrunner,khandelwalPalazzoBrunner}
as thermodynamical resources
have been recently studied, in our opinion,
the answer to the above
questions is still inconclusive.
Here we show that  quantum entanglement plays a central role
in thermodynamics:
we prove that when work is performed on a system
$\sys$, which is in contact with reservoirs $\env_{R}$ and
$\env_{L}$, time-extensive entanglement between
environmental degrees of freedom is unavoidable
at sufficiently low temperatures.

Hints about the existence of this entanglement,
induced
by the driving, were found in \cite{pazFreitas17,pazFreitas18}
while studying the nature of the heat flow
between the environments. It was shown
that in the stationary regime the energy stored in
each environment
varies due to two processes. First,
the resonant absorption (or emission)
from (or into) the driving field can transport
an excitation in a
mode with frequency $\omega_{i}$
to a mode with frequency $\omega_{i} + k \omega_{d}$,
where $\omega_{d}$ is the driving frequency and $k$ is an integer
number. Naturally, this process is not present at
zero temperature, where
there are no excitations to be transported around. Thus, at very
low temperature, the environmental energy varies
because of a different process:
the energy of the driving is dumped into two
modes whose frequencies satisfy the condition
$\omega_{i} + \omega_{j} = k \omega_{d}$.
In \cite{pazFreitas17,pazFreitas18} this was
interpreted as the non-resonant creation of a pair
of excitations, one in each mode, a process
similar to the one underlying the dynamical Casimir
effect \cite{dodonov,wilsonJohanssonPourkabirian}. However, the arguments in
\cite{pazFreitas17,pazFreitas18}
are a conjecture and
not a proof of the existence of entangled pairs. Here we present a rigorous proof of this fact by showing
that the environmental modes become entangled below a certain temperature. Moreover we show that entanglement can persist for temperatures which are higher than the lowest achievable ones by physically relevant cooling methods (analyzing both the resolved sideband and Doppler limits). Although our proof restricts to linear quantum open systems \cite{martinezPaz} (see below), as the process responsible for the generation of entanglement is the same enforcing the validity of the third law of thermodynamics \cite{pazFreitas17,pazFreitas18}, it is natural to conjecture that our result is not a mere consequence of linearity but a general one.

The paper is organized as follows: In Section 2 we present our model, which is a generalization of the standard quantum Brownian motion (QBM) model including a time-dependant driving field, and present its formal solution. In Section 3 we show how to explicitly solve this model for the case of a periodic driving using Floquet theory. In Section 4 we use the previous results to compute correlation functions between different environmental modes and we discuss their properties in the long time regime. In Section 5 we study the entanglement between environmental modes at zero and non-zero environmental temperature. We provide a simple analitycal expression to compute an entanglement measure (the logarithmic negativity) that shows entanglement is created at a constant speed (time extensivity) in a physically relevant range of parameters. We also provide a simple formula to predict the temperature at which entanglement vanishes due to thermal excitations. We summarize our results in Section 6.

\section{The model}
\begin{figure}[htp]
    \begin{center}
    \includegraphics[scale=.5]{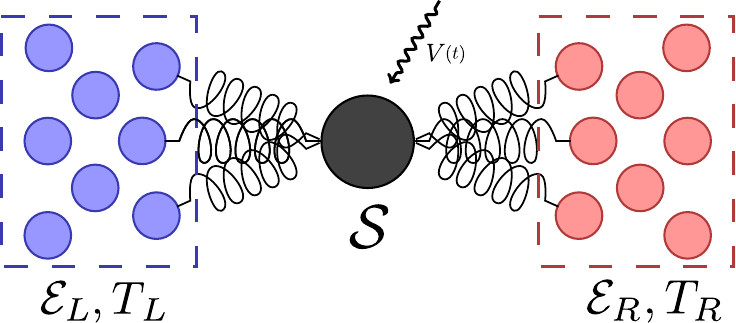}
    \caption{A parametric oscillator $\sys$ driven by $V \left( t \right)$ is coupled with two environments $\env_{L}$ and $\env_{R}$ at temperatures $T_{L}$ and $T_{R}$, respectively.}
    \label{sistema}
\end{center}
\end{figure}

We consider the following generalization of the
quantum Brownian motion model \cite{huPazZhang,caldeiraLeggett}:
a system $\sys$
(a parametric oscillator with
coordinate $x$) is coupled with an environment
$\env$ formed by
independent oscillators
(whose coordinates are
denoted as $q_{i}$, with $i = 1, \dots, N$). The
environment is divided into two pieces
$\env_{\alpha}$, where $\alpha=R,L$,
each of which
consists of the oscillators $q_{i}$ with $i \in \env_\alpha$.
Each
$\env_{\alpha}$ is initially prepared in a thermal state with
temperature
$T_{\alpha}$. This model represents the physical situation shown in Figure \eqref{sistema}.

Thus, our model describes the closed Universe formed by the system $\sys$ and the environment $\env$, whose dynamics is governed by the total Hamiltonian $H_{T} = H_{\sys} + H_{\env}
 + H_{\sys,\env}$. Here, the system's
Hamiltonian is
$H_{\sys}= p^{2}/2m+ m V(t) x^{2}/2$ while the
environmental and interaction terms
are, respectively, $H_{\env} = \sum_{i} \left( p_{i}^{2}/2m_{i} +
m_{i} \omega_{i}^{2} q_{i}^{2}/2 \right)$ and $H_{\sys,\env}=
x \sum_{i} \lambda_{i} q_{i}$. An important
feature of the environment is its spectral
density
$I \left( \omega \right) = \sum_{i} \lambda_{i}^{2} \delta \left( \omega - \omega_{i} \right)
/m_{i} \omega_{i}$
(the spectral density of each
$\env_{\alpha}$ is denoted as $I_{\alpha} \left( \omega \right)$ and
is defined in the same way, restricting
the summation to
$i\in\alpha$). As we are insterested in studying correlations between the environmental modes, we will solve this model in a way which is different from the one used in standard treatments of QBM \cite{huPazZhang}. Thus, we solve the full Heisenberg equations of motion of the coupled system, which read
\begin{equation}
    \begin{dcases}
        \ddot{q}_{i} + \omega_{i}^{2} \, q_{i} = - \lambda_{i} \, x / m_{i} & \quad i = 1, \dots, N \\
		\ddot{x} + V \left( t \right) x = - \textstyle{\sum_{i = 1}^{N}} \lambda_{i} \, q_{i} / m &
    \end{dcases}
    \label{heisenbergEq}
\end{equation}
Interestingly the above equations can be formally solved and the solutions expressed in terms of two decoupled sets of operators. One of them, which we denote $q_{i}^{h} \left( t \right)$, acts on the environmental state space, and the other one, which we denote $x_{h} \left( t \right)$, acts on the system state space. The operators $q_{i}^{h} \left( t \right)$ are the free Heisenberg operators of the environmental modes 
\begin{equation}
    q_{i}^{h}(t) = q_{i,0} \cos \left( \omega_{i} t \right) + p_{i,0} \sin \left( \omega_{i} t \right)/m_{i} \omega_{i}, 
\end{equation}
where $q_{i,0}$ and $p_{i,0}$ are Schr\"oedinger operators. In turn, $x_{h} \left( t \right)$ is a dressed operator for $\sys$ that satisfies the linear equation
\begin{equation}
    \ddot{x}_{h}  + V_{R} \left( t \right) x_{h} +
\gamma \ast \dot x_h = 0,
    \label{sistemaHomogenea}
\end{equation}
where the notation $ F \ast f = \int_{0}^{t} dt^{\prime} F \left( t, t^{\prime} \right) f \left( t^{\prime} \right)$ is used. Above, $\gamma$ is the dissipation kernel
defined as $\gamma \left( t , t^{\prime} \right) = \int d \omega \, I \left( \omega \right) \cos \left[ \omega \left( t - t^{\prime} \right) \right] / m\omega$, and $V_{R} \left( t \right) = V \left( t \right) - \gamma \left( 0 \right)$ is the renormalized potential. The solution of the full Heisenberg equations \eqref{heisenbergEq} is
\begin{equation}
    \begin{dcases}
    q_{i} & = q_{i}^{h} + K_{ij}^{ \left( 1 \right)} \ast q_j^{h} + K^{\left( 2 \right)}_{i} \ast x_{h},\\
    x & = x_{h} + K^{\left( 3 \right)}_{j} \ast q_{j}^{h},
    \end{dcases}
    \label{evolucionCoordenadas}
\end{equation}
where a summation over repeated indices is implicit. The kernel $K^{\left( 1 \right)}$, which plays
a central role in our calculations, is
\begin{equation}
    K_{i j}^{\left( 1 \right)} \left( t , t^{\prime} \right) = \lambda_i \lambda_j \int_{t^{\prime}}^{t} d \tau \sin \left[ \omega_i \left( t - \tau \right) \right] G \left( \tau , t^{\prime} \right) / m m_{i}\omega_{i},
\end{equation}
where $G$ is the Green function of Eq. \eqref{sistemaHomogenea} above. The other two kernels are
\begin{equation}
    \begin{dcases}
        K_{i}^{\left( 2 \right)} \left( t , t^{\prime} \right) & = - \lambda_{i} \, \text{sin} \left[ \omega_{i} \left( t - t^{\prime} \right) \right] / m_{i} \, \omega_{i} \\
        K_{j}^{\left( 3 \right)} \left( t , t^{\prime} \right) & = - \lambda_{j} \, G \left( t , t^{\prime} \right) / m.
    \end{dcases}
\end{equation}

So far the expressions we obtained for $q (t)$ and $x (t)$ are exact. To compute them explicitly we need to solve the equation for $G$, a task that is feasible for certain spectral densities and for simple forms of $V (t)$ (In fact, once we do this we can express $x_{h} (t)$ as a linear combination of the Schr\"odinger operators $x_{0}$ and $p_{0}$). In the case of constant $V(t)$, the expressions presented in Eq. \eqref{evolucionCoordenadas} can be used to obtain all the known results for the standard QBM model, and also to compute environmental correlations. Below we will discuss the solution for a time-periodic driving field. 

\section{Solution for periodic driving using Floquet theory}
In order to deal with the explicit
time dependence induced by $V \left( t \right)$ we use Floquet theory.
For a periodic driving with
$V \left( t \right) = \sum_{k} V_{k} e^{i k \omega_d t}$, $G \left( t , t^{\prime} \right)$ can always be written as 
\begin{equation}
    G \left( t , t^{\prime} \right) = \textstyle{\sum_{k}} A_{k} \left( t - t^{\prime} \right)  e^{i k \omega_{d} t},
\end{equation}
where $A_{k}$ vanishes for negative arguments. For $G$ to be a Green fuction of Eq. \eqref{sistemaHomogenea}, $A_{k}$ must satisfy a linear set of coupled differential equations. Instead of explicitly writing this system, we will simply present an equivalent one involving their Laplace transform $\tilde{A}_{k} (s) = \int_{0}^{\infty} dt \, A_{k} \left( t \right) \, e^{-st}$:
\begin{equation}
    \tilde{g}^{-1} \left( s + i k \omega_{d} \right) \tilde{A}_{k} \left( s \right) + \textstyle{\sum_{n \neq 0}} V_{n} \, \tilde{A}_{k - n} \left( s \right) = \delta_{k 0}.
    \label{eqAk}
\end{equation}
Here, $\tilde{g}$ is the Laplace transform of the
static Green
function  (obtained when $V \left( t \right) = V_{0}$),
which is
$\tilde{g}^{-1} \left( s \right)
= s^{2} + \omega_{r}^{2} + s \, \tilde{\gamma} \left( s \right)$
(where $\omega_{r}^{2} = V_0 - \gamma \left( 0 \right)$
is the renormalized frequency and $\tilde{\gamma}$ is the Laplace transform of $\gamma$). The above system of equations can be simply solved by using a perturbative series expansion which is valid when the Fourier coefficients of the potential $\lvert V_{k} \rvert$ are small. The validity of this approximation also requires the frequency of the driving $\omega_{d}$ to be detuned from the parametric resonance. Thus, the solution of \eqref{eqAk} satisfies the following recurrence relation:
\begin{equation}
    A_{k}^{\left( m \right)} \! \left( s \right) = \tilde{g} \left( s + i k \omega_{d} \right) [ \delta_{k 0} - \textstyle{\sum_{n \neq 0}} V_{n} \, \tilde{A}_{k - n}^{\left( m - 1 \right)} \left( s \right) ],
    \label{potenciasV}
\end{equation}
for $m\geq 1$, with 
$A_{k}^{\left( 0 \right)} \left( s \right) = \tilde{g} \left( s + i k \omega_{d} \right) \, \delta_{k 0}$.
It is interesting to note that the coefficientes $\tilde{A}_{k}$ satisfy certain properties which can be interpreted as a generalization of the static fluctuation-dissipation relation to the driven case. After some algebraic manipulations one can show that
\begin{equation}
	\text{Im} [ \tilde{A}_{0} \left( i \omega \right) ] = - \pi \textstyle{\sum_{k}} I \left( \omega - k \omega_{d} \right) \, \lvert \tilde{A}_{k} \left[ i \left( \omega - k \omega_{d} \right) \right] \rvert^{2} / 2m.
	\label{wkGeneral}
\end{equation}
In the static case, when $V_{k} = 0 \quad \forall \lvert k \rvert \geq 1$ , it reduces to 
\begin{equation}
    \text{Im} [ \tilde{g} \left( i \omega \right) ] = - \pi I \left( \omega \right) \left\lvert \tilde{g} \left( i  \omega \right) \right\rvert^{2} / 2m,
\end{equation}
which is the well known expression for the fluctuation-dissipation relation in the absence of a driving field \cite{huPazZhang}. Bellow we will use the above equations to compute correlations between environmental modes. It is worth noticing that the generalized fluctuaton-dissipation relation considerably simplifies the expressions for the correlators and determines some of their most interesting properties.

\section{Correlation functions between environmental bands}
We use Eq. \eqref{evolucionCoordenadas} to compute all correlation functions between two environmental bands: one of them consists of oscillators whose frequencies are distributed around $\omega_i$ (with a bandwidth $\Delta \omega$) and the other (disjoint) one is centered around $\omega_{j}$. Expressions below will depend on the product $I(\omega)\Delta\omega$ which,
for sufficiently small values of $\Delta\omega$, plays
the role of an effective
coupling strength between the reservoir
band and $\sys$ (since $I(\omega)\Delta\omega\approx\lambda^2/m\omega$). We can show that the environmental correlators can be expressed as the sum of a term that depends on the initial  state of $\sys$  and another one that depends on the
initial correlations within $\env$. In fact, when a stable stationary regime exists, the dependance on the initial conditions for $x$ becomes irrelevant and the dynamics is dominated by the term involving the initial correlations within $\env$. The existence of such regime requires
the energy pumped into $\sys$ to
be dissipated by $\env$, which can be
achieved for small driving amplitudes,
provided that $\omega_d$ is detuned from the
parametric resonance. In Appendix A we include the general form of the equal time correlators $\left\langle \left\{ q_{i} \left( t \right) , q_{j} \left( t \right) \right\} \right\rangle$, both at zero and non-zero environmental temperatures which, together with the expression for the momentum and the cross-correlators, will be used below to compute a measure of entanglement between environmental bands. Here, for the sake of simplicity, we
show just an expression that enables us to discuss a feature that is common to
all correlators which in a physically relevant regime become time-extensive. Thus, we compute the energy $E_{i}$ stored in the band $\omega_i$, with $i\in\alpha$, which is the sum of the two diagonal correlators. It has the following form in the long time limit:
\begin{equation}
    E_i(t)\rightarrow \left[ 1/2 + n_{\alpha} \left( \omega_{i} \right) \right] \omega_{i} + \dot{\mathcal{Q}}_{i} \times t,
    \label{energiaEi}
\end{equation}
with
\begin{equation}
	\begin{aligned}
		&\frac{\dot{\mathcal{Q}}_{i}}{\Delta\omega}=
 		 \textstyle{\sum_{k}} \textstyle{\sum_{\alpha^{\prime}}} \omega_i \left\{ \Theta \left( \omega_{i, k} \right) p_{\alpha,\alpha^{\prime}}^{(k)} \left( \omega_{i} \right) \left[ n_{\alpha^{\prime}} \left( \omega_{i,k} \right) - n_{\alpha} \left( \omega_{i} \right) \right] \right. \\
        & \left. + \Theta \left(-\omega_{i, k} \right) p_{\alpha,\alpha^{\prime}}^{(k)} \left( \omega_{i} \right) \left[ n_{\alpha^{\prime}} \left( \left\lvert \omega_{i,k} \right\rvert \right) + n_{\alpha} \left( \omega_i \right) + 1 \right] \right\},
    \end{aligned}
    \label{heatCurrent}
\end{equation}
where $\omega_{i,k} = \omega_{i} - k \omega_{d}$, $\Theta$ is the step
function, $n_{\alpha} \left( \omega \right)$ is the Planck distribution
with temperature $T_{\alpha}$, and
$p_{\alpha,\alpha^{\prime}}^{(k)} \left( \omega_{i} \right) = \pi \,  I_{\alpha} \left( \omega_i \right)
I_{\alpha^{\prime}} \left( \left\lvert \omega_{i, k} \right\rvert \right) \lvert \tilde{A}_{k}
\left( i \omega_{i,k} \right) \rvert^{2}/2m^2$
is the probability that
mode $\omega_{i}$ in $\env_{\alpha}$ interacts
with mode $\omega_{i, k}$
in $\env_{\alpha^{\prime}}$ through $\sys$. Remarkably,
$E_i$ is a linear function of time, a feature
that is common to all
correlators. This is really
a consequence of the continuum hypothesis for $\env$
since any discrete environment has
recurrence times $t_{rec}\approx 1/\delta\omega$, where
$\delta\omega$ is the smallest frequency splitting in
$\env$. Therefore, although the above expression is valid for
long times, it is not valid for arbitrarily long times
(the continuum hypothesis for arbitrarily
long times, which is equivalent to assuming
an infinite heat capacity for each band,
yields non-physical covariance matrices).
Technically, time extensivity
is obtained in two steps: (i) we write
all second order correlators
as frequency integrals (using Eq.
\eqref{evolucionCoordenadas} and
transforming the summations over discrete
modes into integrals over $\omega$,
weighted by $I(\omega)$);
(ii) we collapse the frequency integrals using
the fact that terms involving
the kernels $K_{ij}^{(1)}$ become
highly peaked functions of $\omega$ in the long time
limit (that can be treated as
Dirac $\delta$-functions
until the time for which the discreteness of the
environmental spectrum becomes relevant).

\begin{figure}[htp]
    \begin{center}
    \includegraphics[scale=.56]{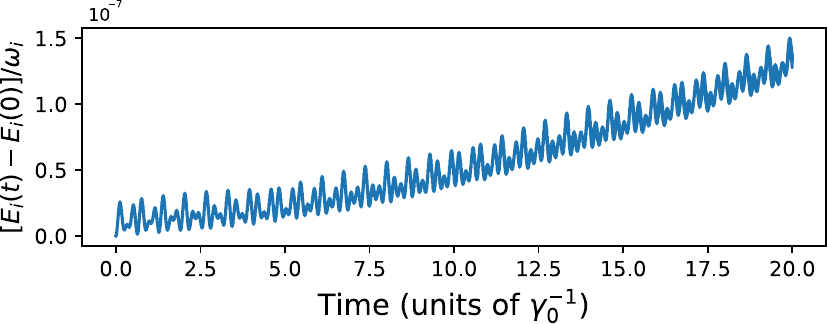}
    \caption{Energy of the mode $\omega_{i}$ as a function of time at zero temperature, for $\omega_{i} = \omega_{d} - \delta$. Here, $\omega_{d} = \omega_{r} - \delta$ with $\delta = 10\gamma_0$, $\omega_{r} = 800\gamma_0$, $V=\omega_r^2/32$, $m=10m_i$ $t=20\gamma_0$ and
$\gamma_0 = 0.005$.}
    \label{energy}
\end{center}
\end{figure}

In Eq. \eqref{heatCurrent} we can see the
hints pointing towards the existence of
entanglement that
motivated our current study. The first term in its r.h.s.
involves the interaction
of a mode $\omega_{i}$ and
another one with frequency $\omega_{i} - k \omega_{d}$,
and describes resonant transport (which can induce
either heating or cooling of the modes,
depending on
the relation between
their populations).
To the contrary, the second term, which
being positive is always associated with
heating, involves the interaction
between modes whose frequencies add up to a
multiple of $\omega_d$. This spliting of the driving
energy between two modes motivated us to ask if
entanglement is generated between them.
We will show below that this is
indeed the case. Before that, let us mention that
our exact formula (which takes into
account the initial transient and does not involve
any time averaging) can be used to obtain
Fig. \eqref{energy}, where
we show the behavior of $E_{i}(t)$
for an Ohmic spectral density
$I(\omega)= 2 m\gamma_0\omega \Lambda^2/\pi (\omega^2+\Lambda^2)$
and a harmonic driving $V_R(t) = \omega_r^2 + V \cos(\omega_d t)$.
The plot (that corresponds to the case where $T_{R} = T_{L} = 0$) shows
a linear behavior for long times, with a
modulation with frequency $\omega_d$.

In the next section we will discuss entanglement between environmental modes focusing on the long (but not infinitly long) time limit, where time extensivity dominates.

\section{Entanglement between environmental bands}
Here we will study entanglement between two environmental bands which are respectively centered around frequencies $\omega_{i}$ and $\omega_{j}$, with $i \in \env_R$ and $j\in \env_L$. For this we will compute a standard measure of entanglement, which is obtained from the reduced density matrix of the two modes $i$ and $j$. We will use the logaritmic negativity $E_{\mathcal{N}} \left( t \right)$ which is a good measure of entanglement since the two-mode state in the stationary regime is Gaussian \cite{adessoIlluminati}. $E_{\mathcal{N}} \left( t \right)$ is obtained, as it is well known \cite{adessoIlluminati,serafiniIlluminatiSiena}, from the smallest symplectic eigenvalue of the covariance matrix corresponding to the partially transposed reduced density matrix of $i$ and $j$. We will obtain our results using two complementary methods. On the one hand, we will derive analytic expressions for $E_{\mathcal{N}} \left( t \right)$, which are valid in the limit of small driving and weak coupling. On the other hand, we will numerically evaluate our exact analytic expressions and compare them with the above aproximation. In the following subsections we will first discuss the case where the environmental temperatures vanish and show that entanglement exists only for modes satisfying the condition $\omega_{i} + \omega_{j} = \omega_{d}$. Then we will present the generalization of this result for the non-zero temperature case. In both cases we will analyze in detail the nature of the results for physically relevant ranges of parameters ($\omega_{d}$, $\omega_{r}$ and $\gamma_{0}$) describing  the most interesting cases that correspond to the resolved sideband and Doppler limits (see below). In the last subsection we present a formula that allows us to compute the maximum temperature at which entanglement between the environmental bands is still present.

\subsection{Entanglement at zero temperature}
We compute the long time limit of the average value of $E_{\mathcal{N}} \left( t \right)$
between two bands whose frequencies add
up to a multiple of the driving frequency $\omega_{d}$ (where the average is taken over a driving cycle). Using the above expressions we show that
\begin{equation}
    E_{\mathcal N} \left( t \right)=\Gamma_0 \times t, 
\end{equation}
where
\begin{equation}
\Gamma_0 = \frac{1}{m} \Delta\omega
	 \left\lvert V \right\rvert \sqrt{I_{R} (\omega_i)  I_{L}(\omega_j)}
	 \left\lvert \text{Re} \left[ \tilde{g}_{i}
	  \, \tilde{g}^{\ast}_{j} \right] \right\rvert,
	\label{casoSencillo}
\end{equation}
with $\tilde g_{l}=\tilde g \left( i \omega_{l} \right)$.
This formula, which is valid to leading order
in $V$ in the weak damping regime,
shows that the driving creates entanglement at a
constant speed. This is one of the central results of this paper.
Instead, if $\omega_i+\omega_j=
\omega_d+\epsilon$ with $\epsilon \ll \omega_{d}$, we find that
$E_{\mathcal N} \left( t \right) \propto  \left\lvert \text{sinc} \left( \epsilon t \right) \right\rvert \, t$,
which is a rapidly decaying function of
$\epsilon$ whose amplitude does not grow with
time. It is interesting to compare the time dependence of
the entanglement with the one of the energy. Although they
are both linear in time they display significant differences.
Thus, in the same regime described above,
Eq. \eqref{energiaEi} can be rewritten as
$E_{i} \left( t \right)  = \omega_{i}/2 +
 \dot{\mathcal Q}_i\times t$ where $\dot{\mathcal Q}_i=
 \pi \omega_{i} \Delta\omega
 \left\lvert V \right\rvert^{2}
 I_{R} \left( \omega_{i} \right)
I \left(\omega_j\right)
\lvert \tilde{g}_{i} \, \tilde{g}^{\ast}_{j} \rvert^{2}/2m^{2}$.  
Notably, the speed
with which the energy grows is of second order
in $V$. Instead, as the entanglement grows when
cross-correlations develop, $\Gamma_0$
turns out to be of first order in the driving amplitude. It is worth mentioning that $E_{\mathcal{N}} \left( t \right)$ is proportional to the environmental bandwith $\Delta \omega$. This is a natural result since entanglement is stablished between modes satisfying the matching condition $\omega_{i} + \omega_{j} = \omega_{d}$. Therefore, $E_{\mathcal{N}} \left( t \right)$ should be proportional to the number of entangled pairs which is linear in $\Delta \omega$.

\begin{figure}[htp]
    \begin{center}
    \includegraphics[scale=.56]{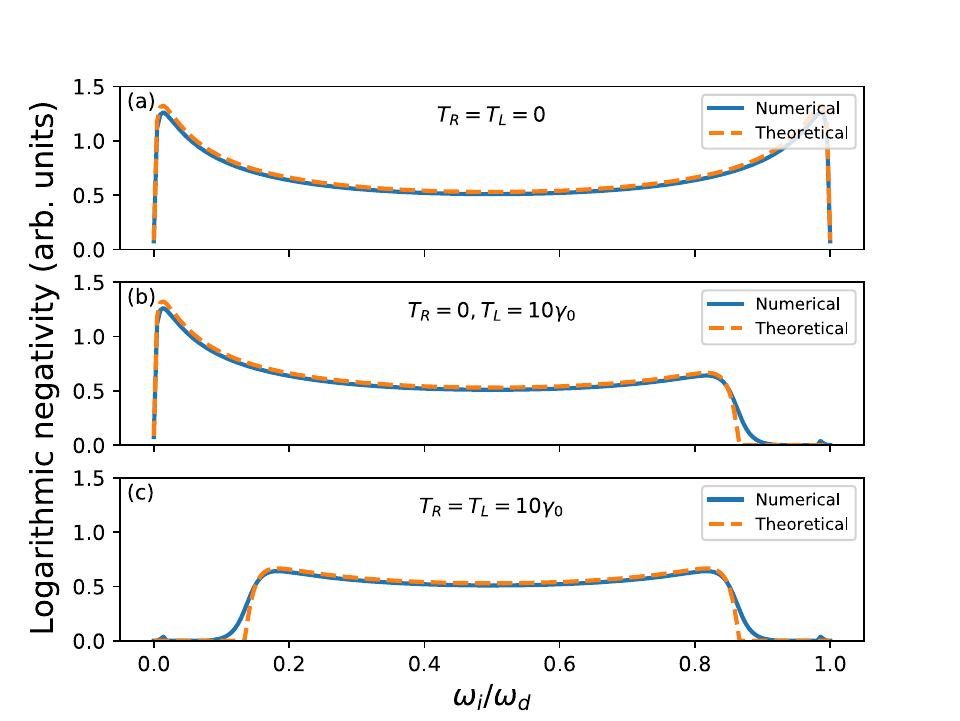}
    \caption{Dependence of the logarithmic negativity $E_{\mathcal N}/E_0$ on the frequency $\omega_{i}$ in the resolved side-band limit when $\omega_{i} + \omega_{j} = \omega_{d}$. (a): When both environments are at zero temperature two peaks and a broadband plateau are
clearly visible. (b): When $T_L$ is
raised to $T_L=10 \gamma_{0}$ the peak at
the right of the figure
(that couples with low frequency
oscillators in $\env_L$) dissappears.
(c): When $T_R = T_L = 10 \gamma_{0}$ only
the central plateau is left. In all cases the
exact result -blue line- coincides with the
analytical estimate given in \eqref{casoSencillo} and \eqref{entrelazamientoTemperatura}. We used the parameters
$\omega_{d} = \omega_{r} - \delta$ with $\delta = 10\gamma_0$, $\omega_{r} = 800\gamma_0$, $V=\omega_r^2/32$, $m=10m_i$, $t=20/\gamma$,
and  $\gamma_0 = 0.005$.}
    \label{entrelazamiento}
    \end{center}
\end{figure}

The amount of entanglement
produced at a certain time
depends on the
frequencies of both modes. By
analyzing the dependence of the entanglement (as measured by the logarithmic negativity) as a function of ($\omega_{i} , \omega_{j}$) we
 can understand the relevance of
our result in the various
regimes defined by
the relation between
$\omega_d$, $\omega_r$ and $\gamma_0$. For example, if
$\omega_d=\omega_r-\delta$ with $\delta\gg\gamma_0$
a maximum of entanglement
is expected for $\omega_i$ close to
$\omega_r-\omega_d = \delta$. This is because oscillators in such band are in the
``resolved sideband'' limit as the decay rate is $\gamma_0\ll\omega_i$
($\omega_r\gg\gamma_0$ is implicit).
In this regime the dominant process is the creation of a pair
of phonons, one in the mode
$\omega_i$ and another one in the mode
$\omega_j=\omega_d-\omega_i=
\omega_r-2\delta$. This is the pair production mechanism that, as
discussed in \cite{pazFreitas18},
sets the limit for the lowest
achievable temperature in sideband resolved
cooling methods. This
intuitive picture is confirmed by an
exact numerical evaluation
of the entanglement, shown in Fig. \ref{entrelazamiento} (a)
where these spectral peaks can be clearly
seen. Dimensional analysis of Eq.
\eqref{casoSencillo} provides us with a natural unit of entanglement
$E_{0} = \gamma_{0} \Delta\omega V / \omega_{r}^{3}$, which we
use to normalize our plots. The agreement between the exact
numerical evaluation and the above
analytic estimate is remarkable.

In turn, when the value of $\delta$ is reduced so that
$\delta\ll\gamma_0$, $E_{\mathcal{N}}$ has
peaks at a band centered around $\omega_i\approx\gamma_0$, which
is not in the resolved sideband regime
(but, as discussed in
\cite{pazFreitas18}, can be associated with the Doppler limit of
laser cooling). We will present further details of the behavior of
entanglement in other regimes elsewhere
\cite{aguilarPaz} and
focus here on studying the robustness of the entanglement in the non-zero
temperature regime.

\subsection{Entanglement at non-zero temperature}
When the initial environmental temperatures are arbitrary, we can use our expressions to
show that the logarithmic negativity
is the maximum between
zero and
\begin{equation}
    E_{\mathcal N}(t)= - S_{i,j} + \Gamma_{\mathcal N} \times t
\end{equation}
where
\begin{equation}
    \begin{aligned}
        \Gamma_{\mathcal N}& = \Gamma_0\ e^{-2S_{i,j}}\
\frac{\nu_{R,i} + \nu_{L,j}}{2}
\left\lvert
\frac{\nu_{R,i} \, \tilde{g}_{i} \, \tilde{g}^{\ast}_{j} +
\nu_{L,j} \, \tilde{g}^{\ast}_{i} \, \tilde{g}_{j}}
{\tilde{g}_{i} \, \tilde{g}^{\ast}_{j} +
\tilde{g}^{\ast}_{i} \, \tilde{g}_{j}}
 \right\rvert
    \end{aligned}
    \label{entrelazamientoTemperatura}
\end{equation}
and $2S_{i,j} = \text{ln} [ ( \nu_{R,i}^{2} + \nu_{L,j}^{2} ) / 2 ] $,
with $\nu_{\alpha,i} =1+
2 \, n_{\alpha} \left( \omega_{i} \right)$ and $\Gamma_{0}$
defined in \eqref{casoSencillo}.
Therefore, at non-zero temperature the entanglement
is also created
at a
constant speed, which is smaller than
the one corresponding to the zero tempearture
case. Moreover, for any finite temperature
there is a latency time $t_{ent}$
until
entanglement is generated. This is fixed
by $S_{i,j}$ and is simply defined as
$t_{ent}= S_{i,j} / \Gamma_{\mathcal N}$. Also, $S_{i,j}$, which is
independent of the driving field,
establishes a lower
bound to the
ammount of entanglement
that is unavoidably lost due to thermal
fluctuations. Interestingly,
it is fixed by the entropy of
the environmental bands: For example, in
the high temperature limit a lower bound
for $S_{i,j}$ is simply given by the average
von Neumann entropy as
$S_{i,j}
\gtrsim [\text{ln} (n_{R}(\omega_{i})) + \text{ln} (n_{L}(\omega_{j}))]/2$.
Fig. \eqref{entrelazamiento} (b) and (c) shows the dependence of $E_{\mathcal{N}}$ on frequency at non-zero temperature in the resolved sideband regime. Again, the agreement between the analytical estimate and the exact numerical result is quite remarkable.

\subsection{Entanglement-breaking temperature}
Entanglement can persist
for a relevant range of temperatures. For
example, when $T_R=T_L$, (the case shown
in Fig. \eqref{entrelazamiento} (c)), the
entanglement persists up to temperatures which
are $15$ times higher than the lowest
temperature  that can be achieved using
sideband resolved cooling methods. In
fact, such methods can cool the band
$\omega_i$ up to an ocupation
number $n_{min}\approx (\gamma_0/2\omega_i)^2$ which, in our
case corresponds to a temperature
$T_{min}\approx 1.7\gamma_0$ but we can show that entanglement
persists up to $T \simeq 30 \gamma_0$.
On the other hand, when only the
temperature $T_R$ is zero,
the entanglement can
persist even for higher values
of $T_L$. Thus, in this case (studied in
Fig. \eqref{entrelazamiento} (b), with
$T_L = 10 \gamma_0$) we found that the
entanglement can persist for temperatures
up to $T\approx 70\gamma_{0}$ which are $40$
times higher than the lowest achievable one
(notice that, in this case,
the entanglement is lost
first for high frequencies in $\env_{R}$,
which correspond to low frequencies in $\env_{L}$,
as expected). It is possible to derive an analytic estimate
for the temperature above which entanglement
vanishes (the proof is shown in Appendix B). Thus, it turns out that
entanglement persists only if
\begin{equation}
    n_{\alpha} \left( \omega_{i} \right) < \frac{2}{\pi} \frac{\gamma_{0}}{2 \omega_{i}} \frac{\Delta \omega}{\omega_{i}}
     \frac{m}{m_{i}}.
    \label{limiteTemperatura}
\end{equation}
This simple formula, notably, accurately predicts
the bounds mentioned above (and
clearly implies that the temperatures above
which
entanglement is lost can be significantly
higher than the minimum cooling temperatures achievable by sideband resolved methods).

\section{Conclusions}
In this paper we studied the creation of quantum correlations between environmental modes in a generalization of the usual QBM model that includes a time-dependent driving enforced on the system $\sys$. The method we used enabled us to solve the model focusing on the behaviour of the environment $\env$ rather than on the one of the driven system $\sys$. Using our analytical expressions we computed a measure of the entanglement between environmental bands, centering our attention in the long-time regime. We showed that with a periodically driven system $\sys$ with a frequency $\omega_{d}$ entanglement between environmental modes satisfying the matching condition $\omega_{i} + \omega_{j} = \omega_{d}$ is always created at low temperatures. This implies that entanglement is in fact unavoidable for driven thermal machines operating at sufficiently low temperatures. We also showed that the entanglement can persist for temperatures  which are significantly higher than the lowest achievable ones for realistic cooling methods. Interestingly, these quantum correlations are created by the same process enforcing the dynamical third law of thermodynamics in linear quantum refrigerators, namely the pair creation induced by the driving. Our results are consistent with recent findings \cite{espositoLindenbergBroeck, ptaszynskiEsposito}  that showed that the entropy production in the stationary regime of non equilibrium systems is dominated by the creation of intra-environment correlations.  Notably, we showed that in a quantum thermodynamical system, such correlations have an intrinsic quantum nature in a relevant regime.

\onecolumngrid
\appendix

\section{Position correlation functions}

Here we will present the general expression for the position correlation function for two modes $i \in \mathcal{E}_{R}$ and $j \in \mathcal{E}_{L}$. In order to obtain the correlation function for just one mode (e.g. $\left\langle \left\{ q_{i} \left( t \right) , q_{i} \left( t \right) \right\} \right\rangle$ instead of $\left\langle \left\{ q_{i} \left( t \right) , q_{j} \left( t \right) \right\} \right\rangle$) we just need to make the replacement $\left\{ i , R \right\} \rightarrow \left\{ j , L \right\}$ and use the generalized fluctuation-dissipation relation when possible. The correlator at zero temperature is
\begin{equation}
	\begin{aligned}
		\left\langle \left\{ q_{i} \left( t \right) ,  q_{j} \left( t \right) \right\} \right\rangle & = \frac{1}{m_{i} \omega_{i}} \, \delta_{i j}\\
		& + \frac{1}{2 m} \frac{1}{\sqrt{m_{i} \omega_{i}}} \frac{1}{\sqrt{m_{j} \omega_{j}}} \Delta \omega \sqrt{I_{R} \left( \omega_{i} \right) I_{L} \left( \omega_{j} \right)} \, \text{Im} \left[ \mathcal{J} \left( \omega_{i} , \omega_{j} , t \right) e^{- i \left( \omega_{i} - \omega_{j} \right) t} - \mathcal{J} \left( \omega_{i} , - \omega_{j} , t \right) e^{- i \left( \omega_{i} + \omega_{j} \right) t} \right] \\
		& + \frac{1}{2 m} \frac{1}{\sqrt{m_{i} \omega_{i}}} \frac{1}{\sqrt{m_{j} \omega_{j}}} \Delta \omega \sqrt{I_{R} \left( \omega_{i} \right) I_{L} \left( \omega_{j} \right)} \, \text{Im} \left[ \mathcal{J} \left( \omega_{j} , \omega_{i} , t \right) e^{ i \left( \omega_{i} - \omega_{j} \right) t} - \mathcal{J} \left( \omega_{j} , - \omega_{i} , t \right) e^{- i \left( \omega_{i} + \omega_{j} \right) t} \right] \\
		& + \frac{1}{\sqrt{m_{i} \omega_{i}}} \frac{1}{\sqrt{m_{j} \omega_{j}}} \, \Delta \omega \sqrt{I_{R} \left( \omega_{i} \right) I_{L} \left( \omega_{j} \right)} \int_{0}^{t} d t_{1} \int_{0}^{t} d t_{2} \, \text{sin} \left[ \omega_{i} \left( t - t_{1} \right) \right] \, \text{sin} \left[ \omega_{j} \left( t - t_{2} \right) \right] \left\langle \left\{ x^{h} \left( t_{1} \right) , x^{h} \left( t_{2} \right) \right\} \right\rangle \\
		& + \frac{1}{4 m^{2}} \frac{1}{\sqrt{m_{i} \omega_{i}}} \frac{1}{\sqrt{m_{j} \omega_{j}}} \Delta \omega \sqrt{I_{R} \left( \omega_{i} \right) I_{L} \left( \omega_{j} \right)} \sum_{\alpha} \int_{0}^{\infty} d \omega \, I_{\alpha} \left( \omega \right) \text{Re} \left[ \mathcal{J} \left( \omega, \omega_{i} , t \right) \mathcal{J}^{\ast} \left( \omega , \omega_{j} , t \right) e^{i \left( \omega_{i} - \omega_{j} \right) t} \right. \\
		& \left. + \mathcal{J} \left( \omega, - \omega_{i} , t \right) \mathcal{J}^{\ast} \left( \omega ,  - \omega_{j} , t \right) e^{- i \left( \omega_{i} - \omega_{j} \right) t} - \mathcal{J} \left( \omega, \omega_{i} , t \right) \mathcal{J}^{\ast} \left( \omega , - \omega_{j} , t \right) e^{i \left( \omega_{i} + \omega_{j} \right) t} \right. \\
		& \left. - \mathcal{J} \left( \omega, - \omega_{i} , t \right) \mathcal{J}^{\ast} \left( \omega , \omega_{j} , t \right) e^{- i \left( \omega_{i} + \omega_{j} \right) t} \right].
	\end{aligned}
\end{equation}
and, at non-zero environmental temperature,
\begin{equation}
	\begin{aligned}
		\left\langle \left\{ \hat{q}_{i} \left( t \right) ,  \hat{q}_{j} \left( t \right) \right\} \right\rangle & = \frac{1}{m_{i} \omega_{i}} \, \left[ 2 \, n_{R} \left( \omega_{i} \right) + 1 \right] \, \delta_{i j}\\
		& + \frac{1}{2 m} \frac{1}{\sqrt{m_{i} \omega_{i}}} \frac{1}{\sqrt{m_{j} \omega_{j}}} \Delta \omega \sqrt{I_{R} \left( \omega_{i} \right) I_{L} \left( \omega_{j} \right)} \, \left[ 2 \, n_{R} \left( \omega_{i} \right) + 1 \right] \\
		& \times \text{Im} \left[ \mathcal{J} \left( \omega_{i} , \omega_{j} , t \right) e^{- i \left( \omega_{i} - \omega_{j} \right) t} - \mathcal{J} \left( \omega_{i} , - \omega_{j} , t \right) e^{- i \left( \omega_{i} + \omega_{j} \right) t} \right] \\
		& + \frac{1}{2 m} \frac{1}{\sqrt{m_{i} \omega_{i}}} \frac{1}{\sqrt{m_{j} \omega_{j}}} \Delta \omega \sqrt{I_{R} \left( \omega_{i} \right) I_{L} \left( \omega_{j} \right)} \, \left[ 2 \, n_{L} \left( \omega_{j} \right) + 1 \right] \\
		& \times \text{Im} \left[ \mathcal{J} \left( \omega_{j} , \omega_{i} , t \right) e^{ i \left( \omega_{i} - \omega_{j} \right) t} - \mathcal{J} \left( \omega_{j} , - \omega_{i} , t \right) e^{- i \left( \omega_{i} + \omega_{j} \right) t} \right] \\
		& + \frac{1}{\sqrt{m_{i} \omega_{i}}} \frac{1}{\sqrt{m_{j} \omega_{j}}} \, \Delta \omega \sqrt{I_{R} \left( \omega_{i} \right) I_{L} \left( \omega_{j} \right)} \int_{0}^{t} d t_{1} \int_{0}^{t} d t_{2} \, \text{sin} \left[ \omega_{i} \left( t - t_{1} \right) \right] \, \text{sin} \left[ \omega_{j} \left( t - t_{2} \right) \right] \left\langle \left\{ x^{h} \left( t_{1} \right) , x^{h} \left( t_{2} \right) \right\} \right\rangle \\
		& + \frac{1}{4 m^{2}} \frac{1}{\sqrt{m_{i} \omega_{i}}} \frac{1}{\sqrt{m_{j} \omega_{j}}} \Delta \omega \sqrt{I_{R} \left( \omega_{i} \right) I_{L} \left( \omega_{j} \right)} \sum_{\alpha} \int_{0}^{\infty} d \omega \, I_{\alpha} \left( \omega \right) \, \left[ 2 \, n_{\alpha} \left( \omega \right) + 1 \right] \\
		& \times \text{Re} \left[ \mathcal{J} \left( \omega, \omega_{i} , t \right) \mathcal{J}^{\ast} \left( \omega , \omega_{j} , t \right) e^{i \left( \omega_{i} - \omega_{j} \right) t} + \mathcal{J} \left( \omega, - \omega_{i} , t \right) \mathcal{J}^{\ast} \left( \omega ,  - \omega_{j} , t \right) e^{- i \left( \omega_{i} - \omega_{j} \right) t} \right. \\
		& \left. - \mathcal{J} \left( \omega, \omega_{i} , t \right) \mathcal{J}^{\ast} \left( \omega , - \omega_{j} , t \right) e^{i \left( \omega_{i} + \omega_{j} \right) t} - \mathcal{J} \left( \omega, - \omega_{i} , t \right) \mathcal{J}^{\ast} \left( \omega , \omega_{j} , t \right) e^{- i \left( \omega_{i} + \omega_{j} \right) t} \right].
	\end{aligned}
\end{equation}
Where the $\mathcal{J}$ function is defined as
\begin{equation}
	\mathcal{J} \left( \omega , \omega_{i} , t \right) = \sum_{k} \int_{0}^{t} d t^{\prime} \, e^{i \left( \omega - \omega_{i} + k \omega_{d} \right) t^{\prime}} \int_{0}^{t^{\prime}} d t^{\prime \prime} \, A_{k} \left( t^{\prime \prime} \right) \, e^{- i \omega t^{\prime \prime}},
\end{equation}
which can be formally solved as
\begin{equation}
	\mathcal{J} \left( \omega , \omega_{i} , t \right) = \sum_{k} \left[ t \, \text{sinc} \left[ \left( \omega - \omega_{i} + k \omega_{d} \right) t / 2 \right] \, a_{k} \left( i \omega \right) \, e^{i \left( \omega - \omega_{i} + k \omega_{d} \right) t / 2} + F_{k} \left( \omega , \omega_{i} \right) \right],
\end{equation}
\noindent where we used
\begin{equation}
	\begin{dcases}
	a_{k} \left( i \omega \right) = \int_{0}^{t} d t^{\prime} \, A_{k} \left( t^{\prime} \right) \, e^{- i \omega t^{\prime}} \\
	F_{k} \left( \omega , \omega_{i} \right) = \frac{a_{k} \left( i \omega \right) - a_{k} \left[ i \left( \omega_{i} - k \omega_{d} \right) \right]}{i \left( \omega - \omega_{i} + k \omega_{d} \right) }
	\end{dcases}
\end{equation}
Note that $F_{k}$ is always finite and $a_{k} \left( i \omega \right) \rightarrow \tilde{A}_{k} \left( i \omega \right)$ in the long time limit.

\section{Proof of the entanglement-breaking temperature formula}
 From now on we will consider two environmental bands $i \in \env_{R}$ and $j \in \env_{L}$ centered around frequencies $\omega_{i}$ and $\omega_{j}$, respectively, such that $\omega_{i} + \omega_{j} = \omega_{d}$. We begin our proof by  defining the dimensionless quantity
\begin{equation}
    \varepsilon_{i} = \frac{\Delta \omega}{m_{i} \omega_{i}^{3}} \, I_{R} \left( \omega_{i} \right)
\end{equation}
for mode $i $ and the analogue $\varepsilon_{j}$ for mode $j $. We will work in the weak coupling regime where $\varepsilon_{i,j} \ll 1$. Using this, and remembering the cannonical form of the two-mode covariance matrix
\begin{equation}
    \sigma = 
    \begin{pmatrix}
        \alpha & \gamma \\
        \gamma^{T} & \beta
    \end{pmatrix},
\end{equation}
the matrices $\alpha$ and $\beta$ can be written as
\begin{equation}
    \begin{dcases}
        \alpha = \nu_{R,i} \, \mathbb{1} /2 + \varepsilon_{i} \, \tilde{\alpha} \\
        \beta = \nu_{L,j} \, \mathbb{1} /2 + \varepsilon_{j} \, \tilde{\beta}
    \end{dcases}
\end{equation}
Now we will make our first approximation, which is $\varepsilon_{i} \simeq \varepsilon_{j}$. Although this is, in fact, an approximation, it is not too far off from reality: $\varepsilon_{i,j}$ is a measure of the coupling of the modes $\left( i, j \right)$ with the system $\sys$, which is roughly the same for entangled $i$ and $j$ modes. From now on we will simply write $\varepsilon$ without a subindex. Under this approximation, we have for all matrices involved,
\begin{equation}
    \begin{dcases}
        \alpha \simeq \nu_{R,i} \, \mathbb{1} /2 + \varepsilon \, \tilde{\alpha} \\
        \beta \simeq \nu_{L,j} \, \mathbb{1} /2 + \varepsilon \, \tilde{\beta} \\
        \gamma \simeq \epsilon \, \tilde{\gamma}
    \end{dcases}
\end{equation}
and
\begin{equation}
    \sigma \simeq \frac{1}{2}
    \begin{pmatrix}
     \nu_{R,i} & 0 \\
     0 & \nu_{L,j}
    \end{pmatrix}
    + \varepsilon
    \begin{pmatrix}
        \tilde{\alpha} & \tilde{\gamma} \\
        \tilde{\gamma}^{T} & \tilde{\beta}
    \end{pmatrix}.
\end{equation}
Using the Caley-Hamilton theorem one can write the determinant of a matrix in terms of traces of powers of such matrix. For example, for $\alpha$ and $\beta$ we have
\begin{equation}
    \begin{dcases}
        \text{det} \left( \alpha \right) \simeq \nu_{R,i}^{2}/4 + \varepsilon \, \nu_{R,i} \, \text{tr} \left( \tilde{\alpha} \right) / 2 + \varepsilon^{2} \, [ \text{tr} \left( \tilde{\alpha} \right)^{2} - \text{tr} \left( \tilde{\alpha}^{2} \right) ] / 2 \\
        \text{det} ( \beta ) \simeq \nu_{L,j}^{2}/4 + \varepsilon \, \nu_{L,j} \, \text{tr} ( \tilde{\beta} ) / 2 + \varepsilon^{2} \, [ \text{tr} ( \tilde{\beta} )^{2} - \text{Tr} ( \tilde{\beta}^{2} ) ] / 2
    \end{dcases}
\end{equation}
In order for this expansion to be exact for $\sigma$, it should be up to order $\mathcal{O} \left( \varepsilon^{4} \right)$. Here is where we will make our second approximation: we will expand $\text{det} ( \sigma )$ up to $\mathcal{O} \left( \varepsilon^{2} \right)$ and assume the next two orders do not significantly contribute to our computations. Thus, we write 
\begin{equation}
	\begin{aligned}
		\det \left( \sigma \right) & \simeq \frac{1}{16} \, \nu_{R,i}^{2} \, \nu_{L,j}^{2} + \frac{1}{8} \varepsilon \, \nu_{R,i} \, \nu_{L,j} [ \nu_{R,i} \, \text{tr} \left( \tilde{\alpha} \right) + \nu_{L,j} \, \text{tr} ( \tilde{\beta} ) ] \\
		& + \frac{1}{8} \varepsilon^{2} \, \nu_{L,j}^{2} \left[ \text{tr} \left( \tilde{\alpha} \right)^{2} - \text{tr} \left( \tilde{\alpha}^{2} \right) \right] + \frac{1}{8} \varepsilon^{2} \, \nu_{R,i}^{2} [ \text{tr} ( \tilde{\beta} )^{2} - \text{tr} ( \tilde{\beta}^{2} ) ] \\
		& + \frac{1}{4} \varepsilon^{2} \, \nu_{R,i} \, \nu_{L,j} [ \text{tr} \left( \tilde{\alpha} \right) \text{tr} ( \tilde{\beta} ) - \text{tr} \left( \tilde{\gamma} \tilde{\gamma}^{T} \right) ].
	\end{aligned}
\end{equation}
For $\gamma$ we make no approximations: $\text{det} \left( \gamma \right) \simeq \varepsilon^{2} \, \text{det} \left( \tilde{\gamma} \right)$. If we analyze the smallest symplectic eigenvalue of the covariance matrix associated with the partially transposed reduced density matrix, we conclude modes $i$ and $j$ are entangled if and only if the following inequality holds true:
\begin{equation}
	\begin{aligned}
		0 & > \frac{1}{2} \left( \nu_{R,i}^{2} - 1 \right) \left( \nu_{L,j}^{2} - 1 \right) + \varepsilon \left( \nu_{R,i} \, \nu_{L,j} - 1 \right) [ \nu_{R,i} \, \text{tr} \left( \tilde{\alpha} \right) + \nu_{L,j} \, \text{tr} ( \tilde{\beta} ) ] \\
		& + \varepsilon^{2} \left( \nu_{L,j}^{2} - 1 \right) [ \text{tr} \left( \tilde{\alpha} \right)^{2} - \text{tr} ( \tilde{\alpha}^{2} ) ] + \varepsilon^{2} \left( \nu_{R,i}^{2} - 1 \right) [ \text{tr} ( \tilde{\beta} )^{2} - \text{tr} ( \tilde{\beta}^{2} ) ] \\
		& + \varepsilon^{2} \, \nu_{R,i} \, \nu_{L,j} [ \text{tr} \left( \tilde{\alpha} \right) \text{tr} ( \tilde{\beta} ) + 2 \, \text{det} \left( \tilde{\gamma} \right) - \text{tr} ( \tilde{\gamma} \tilde{\gamma}^{T} ) ] \\
		& - 2 \varepsilon^{2} \left( \nu_{R,i} \, \nu_{L,j} - 1 \right) \, \text{det} \left( \tilde{\gamma} \right).
	\end{aligned}
\end{equation}
\noindent Now let's suppose that $\nu_{R,i}^{2} - 1 = 2 \, n_{R} \left( \omega_{i} \right)$ and $\nu_{L,j}^{2} - 1 = 2 n_{L} \left( \omega_{j} \right)$ are of order $\varepsilon^{r}$ with $r \in \mathbb{R} > 1$ to be fixed later. If this is the case, then we have
$n_{R,L} \left( \omega_{i,j} \right) \simeq \epsilon^{r}/2 < \epsilon/2$.
Rewriting the last equation, we obtain
\begin{equation}
	\begin{aligned}
		0 & > \varepsilon^{2r} / 2 + \varepsilon^{1 + r} \left( 1 + \varepsilon^{r} / 2 \right) [ \text{tr} \left( \tilde{\alpha} \right) + \text{tr} ( \tilde{\beta} ) ] \\
		& + \varepsilon^{2 + r} [ \text{tr} \left( \tilde{\alpha} \right)^{2} - \text{tr} ( \tilde{\alpha}^{2} ) ] + \varepsilon^{2 + r} [ \text{tr} ( \tilde{\beta} )^{2} - \text{tr} ( \tilde{\beta}^{2} ) ] \\
		& + \varepsilon^{2} \, \left( 1 + \varepsilon^{r} \right) [ \text{tr} \left( \tilde{\alpha} \right) \text{tr} ( \tilde{\beta} ) + 2 \, \text{det} \left( \tilde{\gamma} \right) - \text{tr} ( \tilde{\gamma} \tilde{\gamma}^{T} ) ] \\
		& - 2 \varepsilon^{2 + r} \, \text{det} \left( \tilde{\gamma} \right).
	\end{aligned}
\end{equation}
\noindent Since $r > 1$, we can throw away terms of order $\varepsilon^{2 r + 1}$ and $\varepsilon^{2 + r}$, that are smaller than $\varepsilon^{3}$. Writing $r = 1 + \delta$ we get
\begin{equation}
	\begin{aligned}
		0 & > \varepsilon^{2 \delta} + 2 \varepsilon^{\delta} [ \text{tr} \left( \tilde{\alpha} \right) + \text{tr} ( \tilde{\beta} ) ] \\
		& + 2 [ \text{tr} \left( \tilde{\alpha} \right) \text{tr} ( \tilde{\beta} ) + 2 \, \text{det} \left( \tilde{\gamma} \right) - \text{tr} ( \tilde{\gamma} \tilde{\gamma}^{T} ) ].
	\end{aligned}
	\label{ineq}
\end{equation}
\noindent We want to prove that there is at least one pair of bands $\left( i , j \right)$ that fulfill this inequailty for any $\delta > 0$. Since bands $i$ and $j$ are entangled at zero environmental temperature (because $\omega_{i} + \omega_{j} = \omega_{d}$), we know that $0 > \text{tr} \left( \tilde{\alpha} \right) \text{tr} ( \tilde{\beta} ) + 2 \, \text{det} \left( \tilde{\gamma} \right) - \text{tr} ( \tilde{\gamma} \tilde{\gamma}^{T} )$ holds (that is the condition for bands to be entangled at zero temperature). To prove \eqref{ineq} still holds with the two extra terms, we will consider the case $\omega_{i} = \omega_{j} = \omega_{d} / 2$. We choose this particular case because, when the environmental temperature is higher than zero, entanglement vanishes from the highest frequencies to the lowest ones. Therefore, the entanglement between these modes ($\omega_{i} = \omega_{j} = \omega_{d} / 2$) is the last one to dissapear and this will give us an upper bound for the temperature at which entanglement vanishes for all bands. Thus, in this case, we have $\tilde{\alpha} = \tilde{\beta} = \tilde{\gamma}$ and \eqref{ineq} becomes
\begin{equation}
    0 > \varepsilon^{2 \delta} + 4 \, \varepsilon^{\delta} \text{tr} \left( \tilde{\alpha} \right) + 8 \, \text{det} \left( \tilde{\alpha} \right).
\end{equation}
\noindent For long enough times, $\text{tr} \left( \tilde{\alpha} \right) \sim \varepsilon t / \Delta \omega$ and $\text{det} \left( \tilde{\alpha} \right) \sim - t^{2}$, and therefore the inequality is valid. In short, we proved that if the temperature $T_{R}$ is such that
\begin{equation}
    \boxed{n_{R} \left( \omega_{i} \right) < \frac{1}{2} \varepsilon = \frac{1}{2} \frac{\Delta \omega}{m_{i} \omega_{i}^{3}} I_{R} \left( \omega_{i} \right)}
\end{equation}
\noindent and the analogue for $n_{L} \left( \omega_{j} \right)$ and $T_{L}$, then those modes are entangled for long enough times.

\end{document}